\documentstyle[epsfig,conf_iap,]{article}
\begin{document}
\heading{
Dark Energy: Is it $Q$ or $\Lambda$ ?
} 
\par\medskip\noindent
\author{%
A. Melchiorri$^{1}$, C. J. \"Odman$^{2}$
}
\address{
Astrophysics, Denys Wilkinson Building, University of Oxford, Keble road, OX1 3RH, Oxford, UK\\
$^{2}$  Astrophysics, Cavendish Laboratory, Cambridge University, Cambridge, U.K.
}

\begin{abstract}
New observations of Cosmic Microwave Background Anisotropies, Supernovae
luminosity distances and Galaxy Clustering are sharpening our knowledge 
about dark energy. Here we present the latest constraints.
\end{abstract}
\section{Introduction}

The discovery of a possible accelerated expansion of our present 
universe from type Ia supernovae (\cite{super1}), 
is perhaps the most remarkable cosmological finding of recent years.
Furthermore, the flatness of the universe ($\Omega_{tot}=1$) 
determined by Cosmic Microwave Background observations (\cite{flat}),
togheter with the low matter density ($\Omega_{matter}<0.4$) 
inferred from Large Scale Structure 
(\cite{lowmat}) are suggesting the presence of
a cosmological constant with high statistical significance.
However, the CMB+LSS result relies on the assumption of 
a particular class of models, based on
adiabatic primordial fluctuations, cold dark matter
and a cosmological constant as dark energy component .
In the following we will refer to this class of model as
$\Lambda$-Cold Dark Matter ($\Lambda$-CDM).

This weak point, shared by most of the current studies, 
should not be overlooked: it might be possible that a
different solution to the dark energy scenario than a
cosmological constant can affect the CMB+LSS determination.
It is therefore timely to investigate if the actual CMB 
data is in complete agreement with the $\Lambda$-CDM scenario
or if we are losing relevant scientific informations 
by restricting the current analysis to a subset
of models.

Here first we check to what extent 
modifications to the standard $\Lambda$-CDM scenario
are {\it needed} by current CMB observations with a model-independent 
analysis obtained fitting the actual data with a phenomenological 
function and characterizing the observed multiple peaks
through a Monte Carlo Markov Chain (MCMC) algorithm, 
which allows us to investigate a large number of parameter 
simultaneously ($15$ in our case). 
We found a very good agreement between 
the position, relative amplitude and width of the
peaks obtained through the model independent approach 
with the same features expected in a $4$-parameters 
model template of $\Lambda$-CDM spectra.
Second, since the $\Lambda$-CDM is a good fit to the CMB data, 
we then move to other possible candidates for the dark energy 
component to see what kind of constraint we can obtain.
The common characteristic of alternative scenarios to a cosmological
constant, like quintessence or domain walls, 
is that their equations of state, $w_{dark}=p/\rho$, might 
differ from the value for a cosmological constant, 
$w_{\Lambda}=-1$. Observationally finding $w_{dark}$ different from $-1$ will
therefore be a success for the alternative scenarios.
We found that the present CMB and LSS data can put strong constraints
on the values of $w_{dark}$ assumed as a constant.

\section{Phenomenological fit to the CMB data and agreement with the
$\Lambda$-CDM scenario.}

We model the multiple peaks in the CMB angular spectrum 
by the following function:

\begin{equation}
\ell(\ell+1)C_\ell/2\pi = \sum_{i=1}^N \Delta T_i^2 \exp 
(-(\ell-\ell_i)^2/2\sigma_i^2)
\end{equation}
where, in our case, $N=5$.
We use this formula to make a phenomenological fit
to the current CMB data, constraining the
values of the $15$ parameters $\Delta T_i$, $\ell_i$ and 
$\sigma_i$.
For the CMB data, we use the recent results from the 
BOOMERanG-98, DASI, MAXIMA-1,CBI, and VSA experiments. 
The power spectra from these experiments were estimated in  
$19$, $9$, $13$, $14$ and $10$ bins respectively
(for the CBI, we use the data from the MOSAIC configuration), 
spanning the range $2 \le \ell \le 1500$.
For the DASI, MAXIMA-I and VSA experiments 
we use the publicly available correlation matrices and window functions. 
For the BOOMERanG and CBI experiments we assign a flat interpolation  
for the spectrum in each bin $\ell(\ell+1)C_{\ell}/2\pi=C_B$,  
and we approximate the signal $C_B$ inside 
the bin to be a Gaussian variable.
The likelihood for a given phenomenological model is defined by 
 $-2{\rm ln} {\cal L}=(C_B^{ph}-C_B^{ex})M_{BB'}(C_{B'}^{ph}-C_{B'}^{ex})$ 
where  $M_{BB'}$ is the Gaussian curvature of the likelihood  
matrix at the peak.

We consider $10 \%$, $4 \%$, $5 \%$, $3.5 \%$  and $5 \%$ 
Gaussian distributed  
calibration errors for the BOOMERanG-98, DASI, MAXIMA-1, VSA, and CBI 
experiments respectively and we include the beam uncertainties 
by the analytical marginalization 
method presented in (\cite{sara}).
The phenomenological fit is operated through a MCMC algorithm.
The MCMC approach is to generate a random walk through parameter space that 
converges towards the most likely value of the parameters and samples the 
parameter space following the joint likelihood, the posterior probability 
distribution. A description of the method used can be found in 
\cite{carolina}.
As reported in \cite{carolina} we found an excellent agreement 
between the $\Lambda$-CDM scenario and the model-independent fit.
This result is evident in Figure 1 where
we plot the best-fit phenomenological model with the best-fit
obtained under the assumption of a $4$ parameters $\Lambda$-CDM template.

\begin{figure}[ht]
\centerline{\rotatebox{270}{\psfig{figure=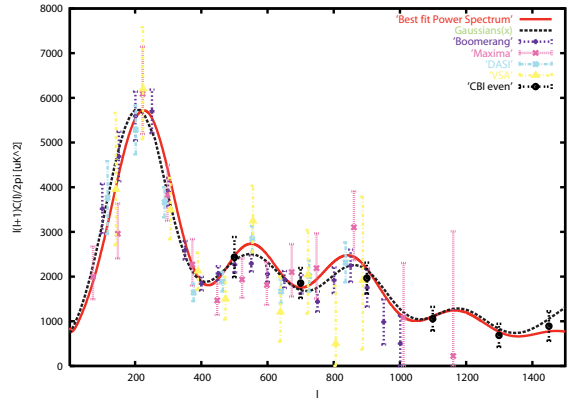,width=5.3cm}}}
\caption{Comparison of the best-fit theoretical
power spectrum and the best-fit model-independent power spectrum.
}
\label{bestfits}
\end{figure}

\section{Quintessence.}

\begin{figure}[ht]
\centerline{\psfig{figure=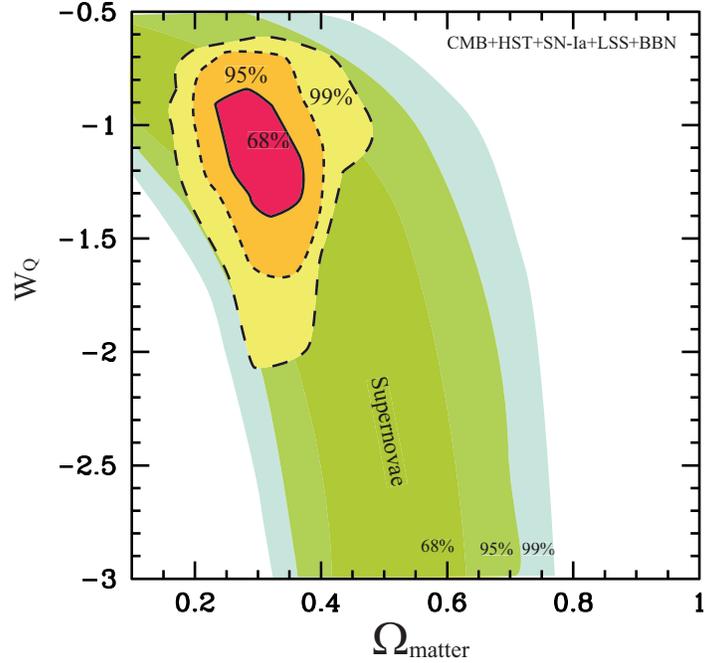,width=9.3cm}}
\caption{Likelihood contours for the energy density and equation
of state of dark energy.Picture taken from \cite{mmot}.}
\label{bestfits}
\end{figure}

Since the present CMB data is in wonderful agreement with
the $\Lambda$-CDM scenario, we can now try to extend the 
set of theoretical models, allowing for values of the
dark energy equation of state different from
$-1$.
However, as already stressed by many authors 
(see e.g. \cite{rachel}) it is necessary to combine the 
CMB data with different data sets in order to obtain
reliable constraints, since each dataset suffers from 
degeneracies with the remaining cosmological parameters. 
Even if one restricts consideration to flat universes 
and to a value of $w_{dark}$ constant
in time then the SN-Ia luminosity distance and position of the
first CMB peak are highly degenerate in $w_{dark}$ and 
$\Omega_{dark}$, the energy density in quintessence.
In Fig. 2 we report the $68 \%$ and $95 \%$ likelihood contours
in the $\Omega_{dark}-w_{dark}$ plane \cite{mmot} with the inclusion of the 
new CBI, VSA and Archeops datasets.

The equation of state parameter is constraied to be
 $-1.62<w_{dark}<-0.76$ at 
$95 \%$ c.l., in agreement with the $w_{dark}=-1$ 
cosmological constant case and giving no support to a 
quintessential field scenario with $w_{dark} > -1$.
A frustrated network of domain walls or
an exponential scaling field are excluded at high 
significance. In addition a number of quintessential models 
are highly disfavored, like, for example, power law potentials with 
$p\ge 1$ . However dark-energy 'phantom' models 
with $w<-1$ (see e.g. \cite{caldwell}) can be in agreement
with the data we considered. 
The result is consistent with other recent independent analyses
(see \cite{ofer} in these proceedings and references therein).

\acknowledgements{We are grateful to Rachel Bean, 
Anthony Lasenby, Laura Mersini, Mike Hobson, and Mark Trodden.}

\vfill

\begin{iapbib}{99}{

\bibitem{super1}  P.M. Garnavich et al, Ap.J. Letters \textbf{493}, L53-57
(1998); S. Perlmutter et al, Ap. J. \textbf{483}, 565 (1997); S.
Perlmutter et al (The Supernova Cosmology Project), Nature \textbf{391} 51
(1998); A.G. Riess et al, Ap. J. \textbf{116}, 1009 (1998);
B.P. Schmidt, Ap. J. \textbf{507}, 46-63 (1998).

\bibitem{flat} C.B. Netterfield et al., [astro-ph/0104460],
C. Pryke et al., [astro-ph/0104489], A. Lee et al., [astro-ph/0104459].

\bibitem{lowmat} G.~Efstathiou {\it et al.}, arXiv:astro-ph/0109152;
N.~A.~Bahcall {\it et al.}, arXiv:astro-ph/0205490.

\bibitem{sara} S.~L.~Bridle, R.~Crittenden, A.~Melchiorri, M.~P.~Hobson,
R.~Kneissl and A.~N.~Lasenby, arXiv:astro-ph/0112114.

\bibitem{carolina}C.~J.~Odman, A.~Melchiorri, M.~P.~Hobson and A.~N.~Lasenby,
arXiv:astro-ph/0207286.

\bibitem{rachel}
R.~Bean and A.~Melchiorri,
Phys.\ Rev.\ D {\bf 65} (2002) 041302
[arXiv:astro-ph/0110472].

\bibitem{mmot}
A.~Melchiorri, L.~Mersini, C.~J.~Odman and M.~Trodden,
arXiv:astro-ph/0211522.


\bibitem{caldwell}
R.~R.~Caldwell,
Phys.\ Lett.\ B {\bf 545} (2002) 23
[arXiv:astro-ph/9908168];

\bibitem{ofer}
O.~Lahav,
arXiv:astro-ph/0212358.

}
\end{iapbib}
\end{document}